\documentstyle[preprint,aps,tighten]{revtex}
\begin{document}
\draft
\title{Potential resonant screening effects on stellar $^{12}$C+$^{12}$C reaction
rates}
\author{R. Cussons, K. Langanke and T. Liolios}
\address{Institute for Physics and Astronomy, University of {\AA}rhus,\\
Denmark}
\date{\today}
\maketitle

\begin{abstract}
The $^{12}$C+$^{12}$C fusion cross sections show resonant behavior  down to
the lowest energies accessible so far in the laboratory. If this tendency
continues into the astrophysical energy range, the stellar $^{12}$C+$^{12}$C
reaction rates have to be corrected for resonant screening effects, in
addition to the conventional screening corrections. We estimate the resonant
screening effects in the weak electron screening limit for hydrostatic
burning and white dwarf environments.
\end{abstract}

\pacs{{\bf PACS.}  26.50.+x, 23.40.-s, 21.60Cs, 21.60Ka}

Nuclear reactions are the energy source for many astrophysical objects \cite
{Rolfs}. When applied in the stellar environment, however, the nuclear
reaction rate, determined from measurements for bare nuclei (for electron
screening corrections to low-energy laboratory cross sections, see \cite
{Assenbaum}), has to be corrected for plasma effects. The medium induces a
screening potential $V_{sc}(r)$ which effectively reduces the Coulomb
barrier between the two colliding nuclei and enhances the reaction cross
section. This effective increase of the in-medium cross section over the
laboratory cross section is conventionally described by an enhancement
factor $f_s$. If $V_{sc}(r)$ is approximately constant during the
penetration process, i.e. for radii smaller than the outer barrier turning
point, the potential can be replaced by the screening energy $U_0=-V_{sc}(0)$%
. Such a situation occurs in astrophysical sites in which the average
Coulomb energy between neighboring ions is much less than the thermal energy
of the plasma. As has been shown by Salpeter \cite{Salpeter54}, in this {\it %
weak screening} limit Debye-H\"uckel theory is applicable leading to an
enhancement of the thermally averaged nuclear reaction rate in the plasma.
In cold ($T \sim 0$), but dense plasma fusion reactions are induced by
density fluctuations. Again, the medium-induced screening potential strongly
enhances the reaction cross sections. The appropriate formalism is developed
for example in \cite{Salpeter69}.

Besides this global screening enhancement an additional plasma correction to
the nuclear reaction rate can occur if the rate is dominated by narrow
resonances. The general theory has been outlined in \cite{Salpeter69}, but
no application has been identified so far. We will point out in this Note
that this situation may occur for the $^{12}$C+$^{12}$C fusion reaction,
where the cross sections show noticeable resonant structures down to the
lowest energies measured so far in the laboratory ($E \sim 2.4$ MeV, \cite
{Kettner}). Until now, the resonant structure of the $^{12}$C+$^{12}$C
fusion cross section has been ignored in astrophysical applications and the
respective reaction rate has been determined as a smooth average over the
resonant contributions. Consistent with this procedure only the global
plasma screening enhancement has been considered. However, if the resonant
structure continues to even lower energies and the astrophysical reaction
rate is due to the contributions of narrow resonances, one then has to
consider that the entrance channel width of these resonances will be
modified in the plasma. This resonant screening effect will reduce the
reaction rate. Our discussion is based on the fact that for $^{12}$C+$^{12}$%
C resonances far below the height of the Coulomb barrier, the entrance
channel width ($\Gamma_a$) is much smaller than the total resonance width.
The latter (which is of order $\sim 100$ keV for the observed resonances
above 2.4 MeV) is also noticeably smaller than the resonance energy. Hence
we can treat the contribution of these resonances to the thermally averaged
reaction rate by the formalism developed for narrow, isolated resonances
(e.g. \cite{Rolfs}), which for the $^{12}$C+$^{12}$C system is: 
\begin{equation}
\langle \sigma v \rangle _{R} =\left( \frac{2\pi }{\mu kT} \right)
^{3/2}\hbar ^{2} 2 (2J+1) \Gamma_a (E_r) \exp \left( - \frac{E_{r}}{kT}%
\right)  \label{eq:trr}
\end{equation}
where $\mu$ is the reduced mass and $E_r$ and $J$ are the energy and total
angular momentum of the resonance.

In the medium the Coulomb potential $V_c(r)$ has to be replaced by $%
V_c(r)+V_{sc}(r)$ ($V_{sc}(r) < 0$). This modification will change the
resonance energy in the plasma to $E_r^\prime < E_r$. Relatedly the
resonance width in the plasma $\Gamma^\prime (E_r^\prime)$ is proportional
to the penetration factor \cite{Clayton} 
\begin{equation}
\Gamma^\prime (E_r^\prime) \sim exp \left( - 2 \sqrt{\frac{2 \mu}{\hbar^2}}
\int_{r_{in}}^{r_{out}} \sqrt{(V_c(r) + V_{sc}(r) -E_r^\prime )} dr \right)
\end{equation}
where $r_{in}$ and $r_{out}$ are the inner and outer barrier turning points,
respectively, and $\mu$ the reduced mass of the two colliding nuclei. The
situation simplifies, if one has $V_{sc}(r) = V_{sc}(0) = -U_0$ for radii $r
< r_{in}$. Then, in a good approximation one has $E_r^\prime = E_r -U_0$. In
the weak screening limit one even has $V_{sc}(r) \approx V_{sc}(0)$ for $%
r<r_{out}$, resulting in $\Gamma^\prime (E_r^\prime) = \Gamma (E_r)$ Such a
situation holds in a massive star ($\sim 20 M_\odot$) where carbon core
burning occurs at temperatures around $10^9$ K and for densities of order $%
\rho = 2 \cdot 10^5$ g/cm$^3$ \cite{Rolfs}. Thus no resonant screening
correction for the $^{12}$C+$^{12}$C reaction cross section is expected in
hydrostatic carbon burning.

In passing we mention that many nuclear reactions occuring in explosive
hydrogen burning \cite{Schatz} are dominated by narrow resonances, e.g. the $%
^{15}O(\alpha,\gamma)^{19}Ne$ reaction which is usually believed to be a key
for the break-out of the matter-flow from the hot CNO-cycle in novae or
x-ray bursters. However, in both sites the weak screening conditions
approximately hold and no resonant screening corrections have to be applied.

In general the magnitude of $V_{sc}(r)$ decreases with increasing radius; at
large distances the medium-induced screening potential cancels the Coulomb
repulsion between the colliding nuclei, i.e. $V_c (r) + V_{sc}(r) = 0$ for
large $r$. If $V_{sc}(r)$ already decreases for radii $r$ under the barrier (%
$r_{in} < r < r_{out}$) the width of the barrier to be penetrated in the
medium is longer than the barrier width for bare nuclei. Correspondingly one
has for the resonance width $\Gamma^\prime (E_r^\prime) < \Gamma (E_r)$
which leads to an additional screening correction of the resonant cross
section which we like to denote by $f_r = \frac{\Gamma^\prime (E_r^\prime)}{%
\Gamma (E_r)}$. The total screening correction is then given by $f_{tot} =
f_s * f_r$. Obviously the resonant screening correction reduces the cross
section in the medium, counteracting the overall enhancement by $f_s$.

For reactions in cold, dense plasma Salpeter and van Horn have derived an
appropriate analytical expression for the resonant screening correction.
This description is appropriate for a carbon white dwarf environment with $%
T=5 \cdot 10^7$ K and $\rho =2 \cdot 10^9$ K \cite{Ichimaru}. Following \cite
{Salpeter69} one then finds a resonant screening correction factor for the $%
^{12}$C+$^{12}$C reaction if this reaction indeed proceeds via a narrow
resonance at energy $E_r$: 
\begin{equation}
f_r = exp \left( - \frac{\pi}{\sqrt{\lambda \epsilon}} \left[ \frac{1.22}{%
\epsilon^3} - \frac{3.1}{\epsilon^6} + \frac{75}{\epsilon^9} \right] \right)
\end{equation}
with the abbreviations 
\begin{equation}
\lambda= \frac{1}{432} \left( \frac{\rho_{12}}{1.629}\right)^{1/3}\; ,
\epsilon = \frac{1}{36} \left(\frac{1.629}{\rho_{12}}\right)^{1/3} \frac{E_r%
}{49.6}
\end{equation}

where $\rho_{12}$ measures the density in $10^{12}$ g/cm$^3$ and the
resonance energy $E_r$ is defined in keV. Obviously the correction becomes
more important with increasing density and at lower resonance energies.

Although the $^{12}$C+$^{12}$C resonance energies below $E \le 2.4$ MeV are
currently not known, we will now assume that the resonance structures
observed in the fusion data at $E \ge 2.4$ MeV continue to lower energies.
While this will probably not strongly affect the smooth energy dependence of
the averaged astrophysical S-factor derived from the data at higher energies
and used for the reaction rate in astrophysical applications so far, it
will, however, change the screening corrections to the rate in the
astrophysical environment. We will estimate the reduction of the $^{12}$C+$%
^{12}$C reaction rate due to the resonant screening correction factor $f_r$
for carbon white dwarf environment hypothetically assuming a resonance in
the energy interval $0.4-2$ MeV.

As is shown in the Fig. 1. the resonant screening correction can lead to a
significant reduction of the reaction rate, which amounts to more than 11
orders of magnitude if the resonance energy is as low as $E_{r}=400$ keV.
Such a change might influence the carbon ignition density in white dwarfs.
We note, however, that despite such a potentially large correction global
screening still dominates, which due to Table 1 of \cite{Ichimaru} amounts
to an enhancement of the rate by 12 orders of magnitude. But this estimate
might also change if the short-ranged potential which gives rise to the
resonances is taken into account \cite{Ogata91}.

In summary, the $^{12}$C+$^{12}$C reaction rate in a white dwarf environment
might change significantly, if the reaction is dominated by narrow
resonances at the relevant energies. This can only be established
experimentally. If such a resonance is indeed being observed it is desirable
to study the plasma enhancement factor due to many-nuclear correlations
including a nuclear potential which supports resonances at low $^{12}$C+$%
^{12}$C scattering energies.

Rob Cussons was partly supported by a Leeds-Aarhus student exchange program.
Theodore Liolios acknowledges a Marie-Curie fellowship from the European
Commission for a stay at the University of Aarhus. Our work is supported in
parts by the Danish Research Council.

\hspace{0.5cm}

\end{document}